\documentclass[traditabstract]{aa} % for the abstract without structuration 
\usepackage{graphicx}
\usepackage{natbib}
\usepackage{multirow}
\usepackage{txfonts}
\begin{document}
\title{Methane, ammonia, and their irradiation products at the surface of an
  intermediate-size KBO?}
   
\subtitle{A portrait of Plutino (90482) Orcus}

\titlerunning{A portrait of Plutino (90482) Orcus}

\author{A. Delsanti \inst{1,2} \and F. Merlin\inst{3} \and
  A. Guilbert--Lepoutre\inst{4} \and J. Bauer\inst{5} \and
  B. Yang\inst{6} \and K.J. Meech\inst{6} }

        \institute{Laboratoire d'Astrophysique de Marseille,
          Universit\'e de Provence, CNRS, 38 rue Fr\'ed\'eric Joliot-Curie,
          F-13388 Marseille Cedex 13, FRANCE\\
          \email{Audrey.Delsanti@oamp.fr} \and
          Observatoire de Paris, Site de Meudon, 5 place Jules Janssen, 92190 Meudon, FRANCE\\
          \email{Audrey.Delsanti@obspm.fr} \and
          University of Maryland, Department of Astronomy, College Park MD 20742, USA\\
          \email{merlin@astro.umd.edu} \and
          UCLA, Earth and Space Sciences department, 595 Charles E. Young Drive East, Los Angeles CA 90095, USA\\
          \email{aguilbert@ucla.edu} \and
          Jet Propulsion Laboratory, M/S 183-501, 4800 Oak Grove Drive, Pasadena, CA 91109, USA\\
          \email{bauer@scn.jpl.nasa.gov} \and
          NASA Astrobiology Institute at Manoa, Institute for Astronomy, 2680 Woodlawn Drive, Honolulu, Hawaii 96822-1839, USA\\
          \email{yangbin@ifa.hawaii.edu,meech@ifa.hawaii.edu} }

   \date{Received: February 20, 2010; Accepted: May 31, 2010;}

   \abstract{Orcus is an intermediate-size 1000km-scale Kuiper Belt
     Object (KBO) in 3:2 mean-motion resonance with Neptune, in an
     orbit very similar to that of Pluto. It has a water-ice dominated
     surface with solar-like visible colors. We present visible and
     near-infrared photometry and spectroscopy obtained with the Keck
     10m--telescope (optical) and the Gemini 8m--telescope
     (near-infrared). We confirm the unambiguous detection of
     crystalline water ice as well as absorption in the 2.2$\mu m$
     region. These spectral properties are close to those observed for
     Pluto's larger satellite Charon, and for Plutino (208996)
     2003~AZ$_{84}$. Both in the visible and near-infrared Orcus'
     spectral properties appear to be homogeneous over time (and
     probably rotation) at the resolution available. From Hapke
     radiative transfer models involving intimate mixtures of various
     ices we find for the first time that ammonium (NH$_4^+$) and
     traces of ethane (C$_2$H$_6$), which are most probably solar
     irradiation products of ammonia and methane, and a mixture of
     methane \textit{and} ammonia (diluted or not) are the best
     candidates to improve the description of the data with respect to
     a simple water ice mixture (Haumea type surface). The possible
     more subtle structure of the 2.2$\mu m$ band(s) should be
     investigated thoroughly in the future for Orcus and other
     intermediate size Plutinos to better understand the methane and
     ammonia chemistry at work, if any. We investigated the thermal
     history of Orcus with a new 3D thermal evolution
     model. Simulations over 4.5$\times$10$^9$yrs with an input 10\%
     porosity, bulk composition of 23\% amorphous water ice and 77\%
     dust (mass fraction), and cold accretion show that even with the
     action of long-lived radiogenic elements only, Orcus should have
     a melted core and most probably suffered a cryovolcanic event in
     its history which brought large amounts of crystalline ice to the
     surface. The presence of ammonia in the interior would strengthen
     the melting process. A surface layer of a few hundred meters to a
     few tens of kilometers of amorphous water ice survives, while
     most of the remaining volume underneath contains crystalline ice.
     The crystalline water ice possibly brought to the surface by a
     past cryovolcanic event should still be detectable after several
     billion years despite the irradiation effects, as demonstrated by
     recent laboratory experiments.}

% \abstract{}{}{}{}{} 
 % 5 {} token are mandatory
 

   \maketitle
%
%________________________________________________________________

\section{Introduction}
\label{sec:intro}

In the past decade, several large Kuiper Belt Objects (KBOs) have been
discovered, unveiling the upper part of the size distribution of these
outer Solar System minor bodies. Eris, discovered in 2003
\citep{Brown+05}, is the largest object known to date, and its
diameter in the 2500--3000km range, exceeding that of Pluto, motivated
the discussion and the creation of a new class of objects in 2006: the
dwarf planets. Several other large objects were discovered since 2003
(Sedna, Makemake, Haumea, etc) and focused the interest of scientists:
the largest of them soon revealed a volatile-rich surface, dominated
by methane, as for Pluto.
\\

Orcus was discovered in 2004 as one of the largest known KBOs
\citep{Brown+04IAU} and is a peculiar object from several aspects. It
revolves around the Sun in a Pluto-like orbit, in 3:2 mean motion
resonance with Neptune (with a semi-major axis of 39.16 AU, an
inclination of 20.6$^{\circ}$, an eccentricity 0.23, a perihelion of
30.26 AU and an aphelion at 48.06 AU). It is currently outbound, very
close to aphelion. With its $\sim$950km diameter
\citep{Stansberry+08,Brown+10}, it is now part of the
intermediate-size objects, and is probably in a transition regime with
respect to volatile surface content. Indeed, smaller KBOs with mostly
featureless near-IR reflectance spectra seem to be totally depleted in
volatiles \citep[see][ and references therein]{Guilbert+09}, while the
largest KBOs show evidence for the presence of volatile ices such as
methane and sometimes molecular nitrogen \citep[like Eris, Sedna and
Makemake, cf.][and references
therein]{Merlin+09,Barucci+05,Brown+07}. A simple model of atmospheric
escape of volatiles dedicated to the Kuiper Belt region was computed
by \cite{Schaller+07} and showed that the objects that are too small
and too hot will most probably lose their pristine volatile surface
inventory (such as CO, CH$_4$ and N$_2$) over the age of the Solar
System, while large and cold objects could have retained these
volatiles to the present day. Objects in an intermediate regime could
have lost some volatiles but retained others, following the different
loss rates at work for the various species. According to this model,
Orcus should have lost all its volatile content. However, traces of
methane (to be confirmed) could explain the near-infrared spectrum of
Orcus \citep{Barucci+08}. This object can therefore provide key
constraints on the current surface volatile inventory of
intermediate-size KBOs.
\\

So far, Orcus' rotation properties are not uniquely defined: 0.03 to
0.18 magnitude amplitude variations are measured over a period ranging
from $\sim$7 to 21h \citep[][and references therein]{Duffard+09},
although \cite{Sheppard+07} finds no lightcurve variations within the
photometric uncertainties of his measurements. Such a flat lightcurve
could be diagnostic of a very slow rotation rate, a small peak-to-peak
lightcurve variation or a nearly pole-on geometric aspect. Also, given
the size of Orcus and the relatively slow rotation rate, we can most
probably expect a circular shape from the relaxation of a fluid body
in hydrostatic equilibrium.
\\
%_____________________________________________________________
\begin{table}
\begin{minipage}[t]{\columnwidth}
  \caption{Summarized properties of the Orcus/Vanth binary
    system. References are: (1) \cite{Noll+08}, (2) \cite{Brown+10}, (3):
    \cite{Brown08}.}
\label{tab:vanth}      
\centering           
\renewcommand{\footnoterule}{}  % to avoid a line before footnotes
\begin{tabular}{c l l c} 
\hline          
\hline   
&Parameter             & Value& Reference \\   
\hline
\multirow{2}*{System} &Angular separation    & 0.26$\arcsec$& (1)\\
                      &Total mass            & 6.3$\times 10^{20}$ kg & (2)\\
\hline   
\multirow{3}*{Orbit}  &Semi-major axis       &8980$\pm$20 km           & (2)\\
                      &Period                &9.5d                     & (2)\\
                      &Inclination           &90$^{\circ}$ or 306$^{\circ}$    & (2)\\ 
\hline
\multirow{3}*{Satellite} &Brightness Frac.     & 8$\%$& (3)\\
                     &Mass ratio           &0.5-0.03\footnote{Assuming an albedo ratio of resp. .5 and .1 and equal densities}  & (2)\\
                     &Diameter             &280-640km\footnote{Assuming an albedo ratio of resp. 1. and .5 and equal densities}                &(2)\\
\hline
\end{tabular}
\end{minipage}
\end{table}
%_____________________________________________________________

To date, Orcus is known to have a single relatively large companion,
Vanth, in a close circular orbit (see details in Table
\ref{tab:vanth}). According to \cite{Brown+10}, the satellite is on a
nearly face-on orbit (with respect to the Sun). Again, Orcus seems to
be a unique case in an intermediate regime: larger KBOs have
relatively small satellites in circular orbits, while small KBOs have
very large satellites \citep{Noll+08}. As described by
\cite{Brown+10}, two scenarios could be invoked to explain the
creation of the Orcus binary system which we will summarize below. In
any case, because the orbit is circular, tidal evolution was at work
at some point of the system's history. First, a catastrophic collision
could be at the origin of the formation of Vanth, as hypothesized for
Pluto-Charon \citep{Canup+05}. Because of the large inclination of
Vanth's orbit, Kozai cycling \citep{Perets+09} could also have
occurred. A second scenario would then be an initial capture of Vanth,
followed by large oscillations of eccentricity an inclination (Kozai
cycling), and when the pericenter drops to a low enough value, tidal
evolution can finally lead the orbit to its current circular shape. As
noted by \cite{Brown+10}, a future discovery of any Orcus coplanar
satellite would rule out this possibility (as for the Pluto-Charon
system). Relative color measurements by the same team indicate that
Vanth is significantly redder than Orcus (another unique property
among KBO binaries), a characteristic that is currently difficult to
understand in the framework of a formation by a giant impact.
\\

In this work, we review the existing spectroscopic investigations of
Orcus and objects with similar spectral and physical properties and
search for additional constraints on the presence of volatiles (and
maybe their irradiation products) on an object that might have
retained some of its original content. We first report additional
0.4--2.5$\mu m$ photometry and spectroscopy obtained from the Mauna
Kea large telescopes. We present Hapke radiative transfer modeling of
the spectral data, the testing for the presence of water ice, methane
and ammonia ices (and their irradiation products, ethane and ammonium
resp.) at the surface and discuss the results. Finally, we use a new
3D thermal evolution model dedicated to the interiors of KBOs
\citep{Guilbert+10} to describe Orcus' thermal history over the age of
the Solar System and the observable consequences of this history on
the current surface.

\section{Observations and data reduction}

\subsection{Visible data}
\label{sec:visible}

% - - - - - - - - - - - - - - - - - - - - - - - - - - - - - - - - - -
\begin{table*}
\begin{minipage}[t]{2.\columnwidth}
%\begin{minipage}[t]{\columnwidth}
  \caption{Comparison of the visible spectral slope S (in \%/100 nm) as computed from linear
    regression using the same algorithm on published data and our work over the common spectral range (0.51--0.76 $\mu$m)}
\label{table:slope}      
\centering          
\renewcommand{\footnoterule}{}
\begin{tabular}{l l c c l  c} 
\hline\hline       
Data reference      & UT date  &r(AU)\footnote{Heliocentric distance from JPL Horizons ephemeris} & $\alpha (^{\circ})$\footnote{Phase angle ($^\circ$) from JPL Horizons ephemeris} &Instrument & Computed S (\%/100nm)\\
\hline                    
This paper          &2005-04-09 &47.67 & 0.99 &Keck + LRIS-RB  & $2.01\pm$0.54\\
\cite{Fornasier+09} &2008-02-04 &47.81 & 0.48 &VLT + FORS2     & $1.75\pm$0.57\\
\cite{deBergh+05}   &2004-04-11 &47.62  &1.02  &VLT + FORS2     & $1.42\pm$0.74\\
\cite{Fornasier+04} &2004-02-29 &47.61  &0.46  &TNG + DOLORES   & $1.80\pm$0.67\\
\hline                  
\end{tabular}
\end{minipage}
\end{table*}
% - - - - - - - - - - - - - - - - - - - - - - - - - - - - - - - - - -
% - - - - - - - - - - - - - - - - - - - - - - - - - - - - - - - - - -
\begin{table*}
  \caption{Near-infrared photometry from Gemini+NIRI (this work) and near-infrared published colors from Orcus.}
\label{tab:nirphotom}      
\centering          
\begin{tabular}{l l l l l l l} 
\hline\hline       
UT date         &$J\pm\sigma$  &$H\pm\sigma$  &$Ks\pm\sigma$ &$J-H\pm \sigma$ &$H-K\pm\sigma$ & Reference\\
\hline
2005--02--20	&17.86$\pm$0.08	&17.73$\pm$0.05	&17.82$\pm$0.05	&0.14$\pm$0.10	&--0.09$\pm$0.07 & This paper\\
2005--02--21	&17.82$\pm$0.04	&17.70$\pm$0.04	&17.81$\pm$0.04	&0.12$\pm$0.07	&--0.11$\pm$0.06 & This paper\\
2005--02--22	&17.85$\pm$0.05	&17.74$\pm$0.05	&17.77$\pm$0.06	&0.11$\pm$0.07	&--0.03$\pm$0.08 & This paper\\
Average         &17.84$\pm$0.11 &17.72$\pm$0.09	&17.80$\pm$0.08	&0.12$\pm$0.14	&--0.08$\pm$0.12 & This paper\\
\hline
Aver. 2004-04-11 to 21 &--- &--- &---                                                 &0.13$\pm$0.05          &0.04$\pm$0.050 &\cite{deBergh+05}\\

\hline
\end{tabular}
\end{table*}
% - - - - - - - - - - - - - - - - - - - - - - - - - - - - - - - - - -

%-------------------------------------------------------------
   \begin{figure}
   \centering
   \includegraphics[width=9cm]{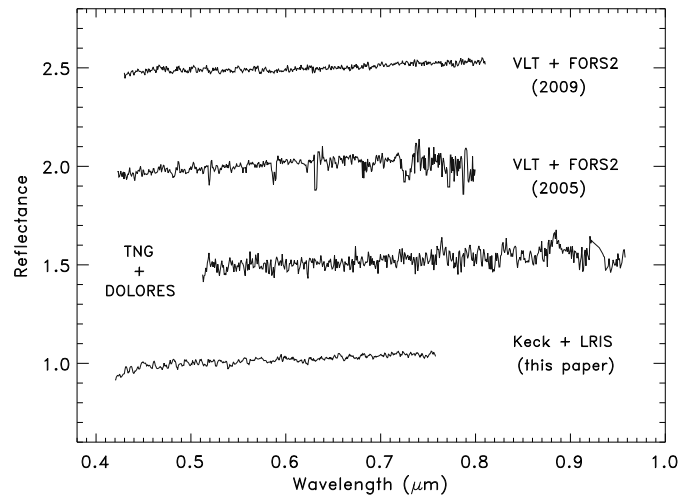}
   \caption{A compilation of available visible spectra of Orcus,
     presented at a resolution of R$\sim$600. From bottom to top: this
     paper, \cite{Fornasier+04}, \cite{deBergh+05},
     \cite{Fornasier+09}. The three top spectra were shifted by 0.5 in
     reflectance for clarity.}
         \label{fig:VisSpec}
   \end{figure}
%
%_____________________________________________________________

   Visible spectra of Orcus were acquired in visitor mode on 2005
   April 9, at the Keck-1 telescope with the LRIS-RB \citep[Low
   Resolution Imaging Spectrometer,][]{Oke+95} instrument mounted at
   the Cassegrain focus. Beam-splitters separate the light between two
   arms, red and blue. The field of view (FOV) of both arms is
   $\sim$7'$\times$6.8'. The blue side is equipped with a mosaic of
   two 2K$\times$4K Marconi (E2V) CCDs with a plate scale of
   0.135''/pixel. The red arm has a Tektronik 2048K$\times$2048K
   detector with a scale of 0.211''/pixel. We used the gratings
   400/3400 on the blue side and 400/8500 on the red side with
   dichroic 560. The resulting wavelength coverage is 0.37--0.83 $\mu
   m$ with a resolution of $\sim$350. We used a long slit of one
   arcsec width that we maintained along the parallactic angle during
   all the observations to minimize the loss of flux due to the
   atmospheric refraction. Arc images were obtained from Hg/Ne/Cd/Zn
   lamp frames on the blue side and Ne/Ar lamp frames on the red
   side. To avoid additional known distortion effects due to the
   telescope flexure, we acquired spectral flat-field and arc frames
   for each science telescope position.
   \\
   We corrected the spectral images for bias, flat-field and cosmic
   rays with the IRAF \citep{Tody86} spectroscopy pack, which we also
   used to calibrate the wavelength. For each instrument arm, we
   extracted the spectra using IRAF/APEXTRACT APALL task and merged
   them into a single 1D spectrum. It was corrected by a solar analog
   (HD 126053) observed close in airmass and time and normalized to
   unity around 0.55 $\mu$m. The resulting 2700s spectrum (which is a
   combination of three individual spectra, 900s each) is presented in
   Fig. \ref{fig:VisSpec}. We do not believe that the slight turn-off
   at 0.42 $\mu$m is real. We inspected the three individual spectra
   and found that although the overall spectral slope remains
   consistent from spectrum to spectrum, small variations were
   observed near the blue end of the individual spectra, which may be
   caused by the poor CCD sensitivity in this wavelength
   region. Therefore, we concluded that the slight turn-down feature
   that appears at 0.42 $\mu$m in the combined spectrum of Orcus is
   not real, but an artifact.
   \\

   We also acquired a series of V band images (through the broadband
   filter centered at 5437$\AA$ and with a FWHM of 922$\AA$) to
   monitor the photometry over the course of the spectroscopy
   observations. Data reduction, photometric calibration and
   measurements were applied following a method described in
   \cite{Delsanti+01}. We found an average value of $V=19.06\pm0.03$,
   which corresponds to $m_V(1,1) = V - 5 \log (r \Delta) =
   2.32\pm0.03$ (with $r$ and $\Delta$ the helio- and geocentric
   distance respectively in AU). \cite{Rabinowitz+07} found a value of
   $H_V = 2.33\pm0.03$ from their solar phase curve study. They
   published a phase correction coefficient for Orcus of
   $0.114\pm0.030$ mag. deg$^{-1}$ in V. As we observed with a phase
   angle of 0.99$^{\circ}$, our phase-corrected absolute magnitude is
   therefore $2.21\pm0.04$.

\subsection{Near-infrared data}
% - - - - - - - - - - - - - - - - - - - - - - - - - - - - - - - - - -
\begin{table}
\begin{minipage}[t]{\columnwidth}
  \caption{List of near-infrared Gemini+NIRI spectroscopy observations.}
\label{tab:specobs}      
\centering          
\renewcommand{\footnoterule}{}  % to avoid a line before footnotes
\begin{tabular}{l l l l l l} 
\hline\hline       
G.\footnote{Grism. J: 1.00--1.36 $\mu$m, H: 1.43--1.96 $\mu$m, K: 1.90--2.49 $\mu$m} & UT start date     & Exp.\footnote{Exposure time in seconds} & Airmass      & Solar An. & Airm.\\
\hline
J     & 2005-02-20 09:43  &1200      & 1.10-1.12	& HIP51125     &1.24\\
H     & 2005-03-01 10:54  &2160      & 1.22-1.41        & HIP47071     &1.24\\
K     & 2005-02-22 09:38  &3600      & 1.09-1.13        & HIP51125     &1.28\\
K     &	2005-02-22 10:52  &2100      & 1.15-1.23        & HIP51125     &1.28\\
\hline                  
\end{tabular}
\end{minipage}
\end{table}
% - - - - - - - - - - - - - - - - - - - - - - - - - - - - - - - - - -

Data were acquired in queue mode at the Gemini North Telescope with
the NIRI (Near InfraRed Imager) instrument on UT 2005 February 20--22
and UT March 1, (see Table \ref{tab:specobs}). The detector is a
1024x1024 ALADDIN InSb array. We used the f/6 camera, which gives a
FOV of 120'$\times$120' and a pixel scale of 0.117 arcsec/pixel in
imaging. Photometry was acquired through the broadband filters J, H, K
short, centered at 1.25, 1.65, 2.15 $\mu$m respectively. We used the
GEMINI NIRI package to produce the master dark and flat-field
frames. To correct for the sky background contribution and recombine
the sub-frames into the final image, we used the IRAF/XDIMSUM
package. We measured the object flux using an aperture correction
method \citep[see][for details on the complete
procedure]{Delsanti+04}. Photometric coefficients (zero-point,
extinction) were obtained from a least-square fit for the standard
star flux measurements. Results are presented in Table
\ref{tab:nirphotom}. Our near-infrared colors are compatible with
those of \cite{deBergh+05}.
\\

We acquired the spectra through the 6--pixel wide slit and the f/6 J
grism (1.00--1.36 $\mu$m useful range with the ``blue'' slit,
resolution of $\sim$480), H grism (1.43--1.96 $\mu$m, R$\sim$520), and
K grism (1.90--2.49 $\mu$m, R$\sim$520). Circumstances are detailed in
Table \ref{tab:specobs}. We used the usual AB nodding technique to
acquire the sub-frames of a cube of data (the telescope was dithered
by $\sim$20'' in between sub-images). Solar analogs were observed to
correct science spectra from the telluric lines and the contribution
from the Sun's spectrum. We used the GEMINI GNIRS package to produce
the ancillary frames (dark, flatfield, arc). Cubes of dithered spectra
were corrected from the dark current, flatfield, and the distortion
effects. The AB pairs were subtracted (A--B and B--A) to correct from
the sky contribution and were re-aligned into a final composite 2D
image. The 1D spectrum was extracted with the IRAF/APEXTRACT APALL
task. Wavelength calibration was performed with IRAF/ONEDSPEC package.

\section{Results}

\subsection{A featureless neutral visible spectrum}

Orcus shows a linear and featureless spectrum (within the noise level)
over the visible range, which is compatible with previously published
data. Figure \ref{fig:VisSpec} shows the spectra available in the
literature, presented at a resolution of R$\sim$600 \citep[data from
this work and][were degraded to match the R$\sim$600 resolution of the
FORS2 data]{Fornasier+04}. We computed a linear regression using the
same algorithm for all published spectra and our work (on the
original, non-degraded data), over the wavelength range that is common
to all spectra (0.51 -- 0.76 $\mu$m) in order to compare the results:
values of the spectral slope in \%/100nm are reported in Table
\ref{table:slope}. They are compatible at one sigma level and are
characteristic of a quasi-neutral surface with solar colors. We do not
believe the slight turn-off at 0.4 micron is real as explained in
Sect. \ref{sec:visible}.  Orcus shows very consistent visible spectral
properties over time, which strongly points to a homogeneous surface,
at least in the optical.

\subsection{Crystalline water ice, ammonia, and methane?}
\label{nir}
%-------------------------------------------------------------
   \begin{figure}
     \centering
     \includegraphics[width=9cm]{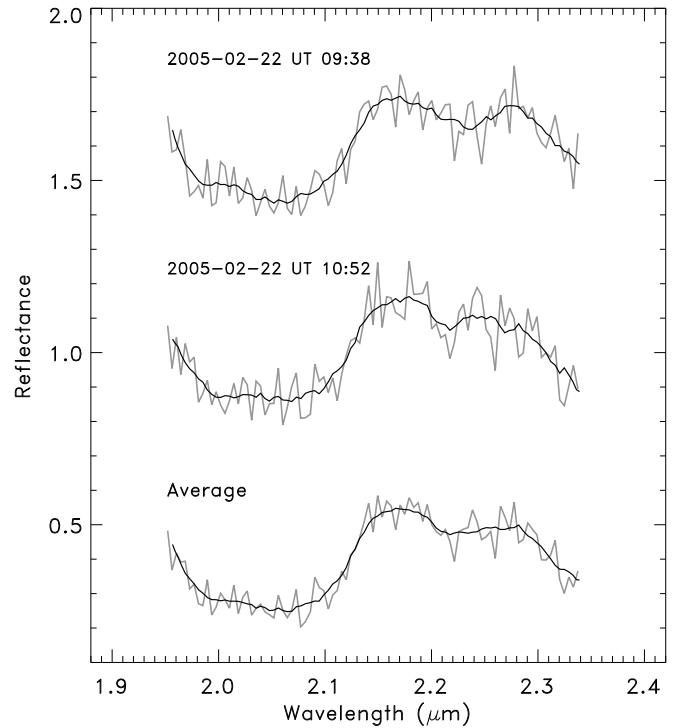}
     \caption{Orcus spectra taken with the K grism on 2005-02-22. Top
       and bottom spectra were shifted by +0.6 and -0.6 resp. in
       reflectance for clarity. The solid line is a running mean to
       guide the eye (smooth box with a width of $\sim55\AA$).}
              \label{fig:Kband}
   \end{figure}
%______________________________________________________________
%_____________________________________________________________
%
\begin{table}
  \caption{Water ice near-infrared absorption band depths and central positions for Orcus.}
\label{tab:prof} 
\centering       
\begin{tabular}{lll}
\hline\hline       
Data   &Central $\lambda$ ($\mu$m) & Depth (\%)\\
\hline                        
H grism             &1.50   &25.5$\pm$3.6\\ 
H grism             &1.65   &13.2$\pm$4.5\\
K grism UT 09:38    &2.05   &28.2$\pm$3.6\\
K grism UT 10:52    &2.05   &22.2$\pm$6.7\\
\hline
\end{tabular}
\end{table}
%
%_____________________________________________________________
%  
%-------------------------------------------------------------
   \begin{figure*}
     \centering
     \includegraphics[width=13cm]{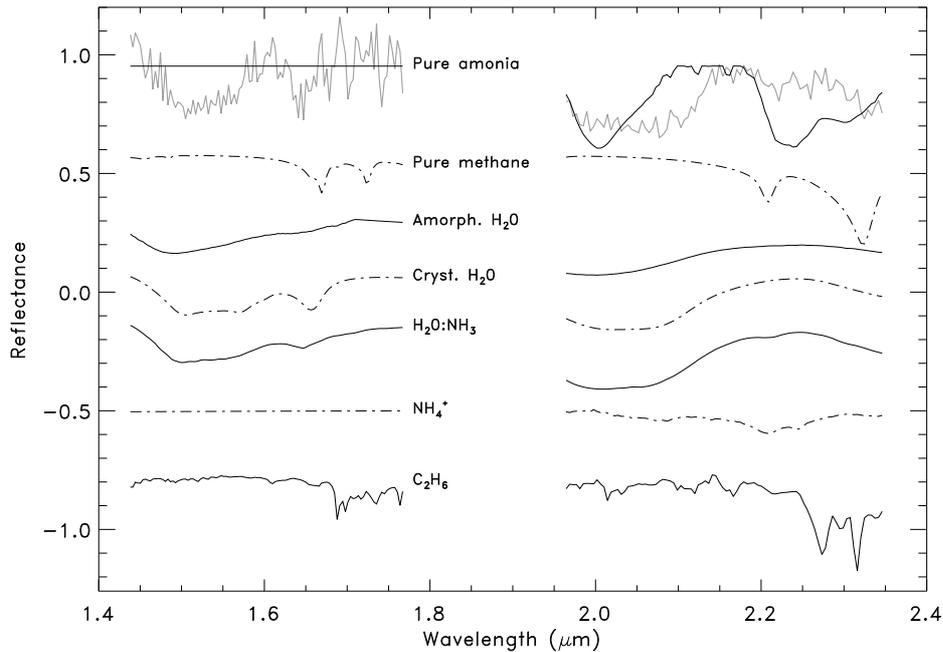}
     \caption{A close up of Orcus spectrum in H and K bands (top, grey
       line) and the reflectance of, from top to bottom: pure ammonia
       and, shifted by several units for clarity, pure methane ice,
       amorphous water ice, crystalline water ice, hydrated ammonia
       (H$_2$O:NH$_3$, 99:1), ammonium and ethane. Synthetic ice
       spectra were obtained using Hapke theory
       \citep{Hapke81,Hapke93} with the optical constants quoted in
       this paper and a particle size of 10 $\mu$m.}
              \label{fig:components}
   \end{figure*}
%______________________________________________________________

   We adjusted the visible and near-infrared parts of the spectrum in
   reflectance using the V magnitude computed in Sect.
   \ref{sec:visible} and near-infrared photometry from 2005 February
   20, (see Table \ref{tab:nirphotom} \textit{i.e.} photometry that
   was observed simultaneously with the J band spectrum). In the K
   band, we obtained two consecutive spectra on 2005 February 22,
   which are plotted at the resolution of the instrument in
   Fig. \ref{fig:Kband} along with the average spectrum. The complete
   reconstructed spectrum at the instrument resolutions from 0.40 to
   2.35 $\mu$m is presented in Fig. \ref{fig:models} (using the
   average of K-band spectral data).
   \\
   
   In the near-infrared, the spectrum of Orcus shows deep absorption
   bands centered at 1.5 $\mu$m, 1.65 $\mu$m, 2.05 $\mu$m (attributed
   to the presence of crystalline water ice) and smaller features
   around 2.2$\mu$m. Measured water ice band depths and positions
   (using a continuum removal plus Gaussian fitting method) are
   reported in Table \ref{tab:prof}. Figure \ref{fig:components} shows
   a close-up of the data in the 1.4--2.4$\mu$m range, the synthetic
   reflectance spectrum of different pure and hydrated ices, and some
   of their irradiation products (with 10 $\mu$m size grains). From
   these, we can unambiguously identify crystalline water ice (with
   the broad 1.5 $\mu$m and 2.0 $\mu$m features and the 1.65 $\mu$m
   absorption band). Methane or ammonia ices may account for the 2.2
   $\mu$m absorption band as surmised also by
   \cite{Barucci+08}. However, hydrated and pure ammonia ices do not
   share the same features around 2.2 $\mu$m. The Orcus spectral shape
   beyond 2.23 $\mu$m may also suggest the presence of methane ice.
   Ethane ice is also surmised, as it is an irradiation product from
   methane \citep{Baratta+03}: both components best described the
   near-infrared spectral behavior of Quaoar
   \citep{SchallerQuaoar+07}, which is very similar to that of
   Orcus. We also show a spectrum of ammonium (NH$_4^+$), as it is an
   irradiation product of ammonia and helped to reproduce the
   2.2$\mu$m features in the spectrum of Pluto's satellite Charon
   \citep{Cook+09}.
   \\

\section{Radiative transfer modeling of  the surface composition}

\subsection{Algorithm and parameters}

To better investigate the surface properties of Orcus, we ran a
radiative transfer model following the theory of
\cite{Hapke81,Hapke93} over the complete wavelength range
available. This algorithm fits a synthetic reflectance spectrum to the
data, using reflectance properties (optical constants) from a
user-defined input list of chemical components, their relative
amounts, and a grain size. When applying this method, one has to keep
in mind that the best-fit synthetic spectrum obtained is not a unique
solution to describe the object's surface composition and that some
parameter values might lead to the best fit, although with no physical
meaning. A detailed description of this model and its limits are
presented in \cite{Merlin+10}.

We chose a symmetry parameter of $v = -0.4$ and a back-scattering
parameter $B = 0.5$. These values are close to those computed by
\cite{Verbiscer+98} to describe the surfaces of the icy satellites of
the giant planets. Finally, we computed the geometric albedo at zero
phase angle from Eq. 44 of \cite{Hapke81}. We used the
Marqvardt--Levenberg algorithm to obtain the lower reduced $\chi^{2}$
value for the fit of the synthetic spectra to our data. Each synthetic
spectrum is obtained using an intimate mixture of different compounds
(from a user-defined input list). The free parameters are the grain
size and the relative amount of each chemical compound. We emphasize
that the results presented here are not unique and only show a
possible surface composition.

\subsection{The choice of the chemical components}
\label{components}

As input to the radiative transfer model, we used optical constants of
several ices at low temperature (close to 40K) that are likely to be
present on the surface of an icy body located at $\sim$50 AU such as:
\begin{itemize}

\item
Water ice \citep{Grundy+98}
\item
Pure methane ice \citep{Quirico+97}
\item
Pure ammonia ice  \citep{Schmitt+98}
\item
Diluted ammonia ice (Ted Roush, personal communication)
\item
Pure methanol ice (Quirico \& Schmitt, personal communication). 
\end{itemize}

We also used optical constants of a dark compound such as amorphous
black carbon \citep{Zubko+96}, which is able to reproduce the low
albedo of the majority of TNOs \citep{Stansberry+08}. Space weathering
is known to be responsible for the alteration of the KBOs surfaces
with time through the action of galactic cosmic rays ions, solar wind
plasma ions and energetic particles,
\citep{Strazzulla+91,Cooper+03}. The effect on ices is documented from
laboratory experiments \citep[for instance][]{Brunetto+06}: the fresh
and bright icy mantle becomes spectrally redder and darker with
irradiation. Without any better constraints on the nature of the
irradiation products present on the surface of these objects, we
decided to use tholins, which reproduce the spectral slope behavior
best in the visible \citep[in our case, Titan tholins were found to be
the more appropriate; this compound is generated from the irradiation
of mixtures of N$_{2}$ and CH$_{4}$,][]{Khare+84}.
\\

Additional spectroscopic observations of Orcus in the near infrared
\citep[H+K band only, from][]{Barucci+08} showed deep absorption
features at 1.5 and 2.0 $\mu$m and shallow bands at 1.65 $\mu$m and
close to 2.2 $\mu$m. The presence of crystalline water ice is evident
in the 1.65 $\mu$m feature. The weak 2.2$\mu$m feature is much more
difficult to interpret (owing to the detection level and multiple
candidate components). These authors tested Hapke models of ammonia
hydrate or pure methane ices (separately): none of these attempts
could satisfactorily reproduce the shape of the 2.2$\mu$m feature,
although ammonia hydrate was the favored explanation. In this work, we
will try to better reproduce the general shape of our spectrum on the
full 0.4-2.4 $\mu$m range with the help of the Hapke modeling of
mixtures of ices listed above, assuming an albedo of 0.2
\citep{Stansberry+08}. Below, we will not discuss in detail the
various relative quantities listed in Table \ref{table:models},
because we consider Hapke model results as only qualitatively
indicative of the possible candidate components. The reasons are that:
(i) synthetic models, even when accurately fitted to the data, do not
provide a unique, definitive solution; (ii) the SNR of the data
currently available for Orcus do not allow this detailed
interpretation; (iii) the optical constants from materials used here
where measured in conditions close to those that should prevail on
Orcus, but some of them are missing and approximations were used.

\subsection{A simple water ice mixture}

%-------------------------------------------------------------
   \begin{figure*}
     \centering    
     \includegraphics[width=15cm]{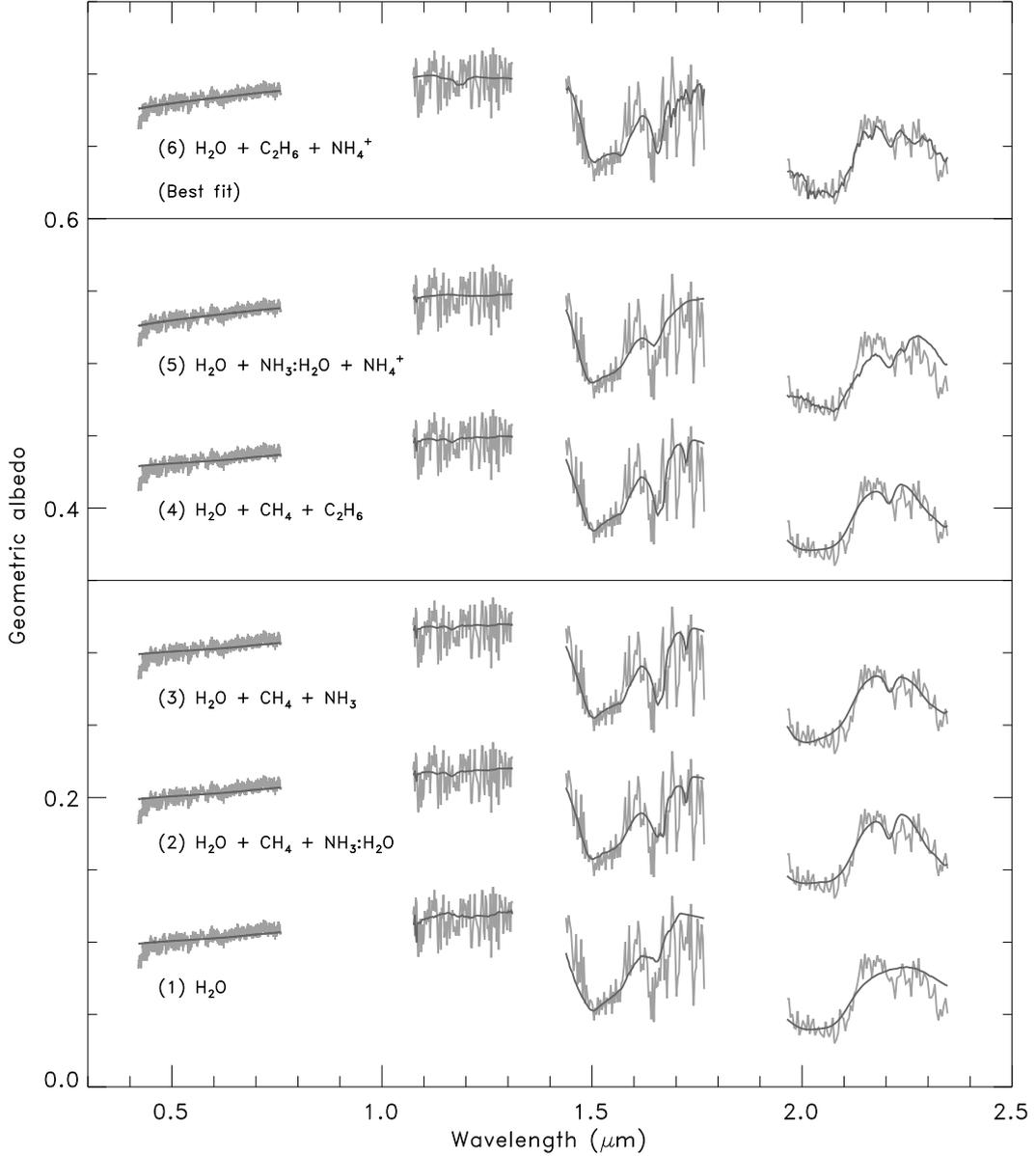}
     \caption{Visible and near-infrared spectrum of Orcus (grey line)
       with synthetic spectra superimposed (solid black line). Spectra
       are shifted in albedo for clarity from -0.1 to about +0.5
       (except model 2). Quantitative description of the models and
       reduced $\chi^2$ are presented in Table \ref{table:models}. The
       bottom spectrum shows that a water ice-only model (Haumea-like
       surface) fails at describing all of Orcus features, especially
       in the K band, although water ice is the likely dominant
       component. The two models above are the best-fit results from
       our next investigation step involving volatiles: they include a
       mixture of methane and ammonia (pure of diluted) in addition to
       water ice (in its amorphous and crystalline state). The middle
       part of the plot illustrates the models including ammonia or
       methane and their respective irradiation products (ammonium and
       ethane). The top part shows our overall best-fit result: a
       mixture of water ice, ammonium (NH$_4^+$), and ethane
       (C$_2$H$_6$).}
     \label{fig:models}
   \end{figure*}
%
%______________________________________________________________
% - - - - - - - - - - - - - - - - - - - - - - - - - - - - - - - - - -
\begin{table*}
  \caption{Quantitative description of the chemical composition
    models with the relative percentage of each component. Grain
    size in micron is indicated in parenthesis. Models are sorted on
    Fig. \ref{fig:models} from bottom (model 1) to top (model 6).}             
\label{table:models}      
\centering          
\begin{tabular}{l | l l l | l l | l | l} 
\hline\hline       
Components                    &Model 1       &Model 2       &Model 3      &Model 4    &Model 5    &Model 6  & Reference\\
\hline                                                                                
Amorphous H$_2$O         &13\% (500)    &5\% (600)    &7\%  (400)   &9\% (370)   &5\% (300)  &1\% (200)   &\cite{Grundy+98}\\
Crystalline H$_2$O       &13\% (40)     &3\% (100)    &13\% (160)   &12\% (150)  &3\% (50)   &19\% (200)  &\cite{Grundy+98}\\
Pure NH$_3$              &---            &---           &2\% (120)    &---          &---         &---          &\cite{Schmitt+98}\\
NH$_3$:H$_2$O (1:99)      &---            &13\% (150)   &---           &---          &11\% (300) &---          &T. Roush (Priv. comm)\\
Pure CH$_4$              &---            &6\% (800)    &5\% (800)    &5\% (800)   &---         &---          &\cite{Quirico+97}\\
Amorphous C              &71\% (10)     &71\% (10)    &71\% (10)    &71\% (10)   &71\% (10)  &58\% (10)   &\cite{Zubko+96}\\
Titan tholin             &3\% (10)      &2\% (10)     &2\% (10)     &3\% (10)    &0\% (10)   &1\% (10)    &\cite{Khare+84}\\
C$_2$H$_6$               &---            &---           &---           &0\% (20)    &---         &7\% (600)   &\cite{Quirico+97}\\
NH$_4^+$                 &---            &---           &---           &---          &9\% (800)  &13\% (800)  &J. Cook (Priv. comm.)\\
\hline
Reduced $\chi^2$         &1.78          &\textbf{1.29}&\textbf{1.32}&1.35 &1.64       &\textbf{1.23}    &\\
\hline                  
\end{tabular}
\end{table*}
% - - - - - - - - - - - - - - - - - - - - - - - - - - - - - - - - - -

Given the obvious features of water ice in its crystalline state in
our data, our first attempt was to use a simple mixture of amorphous
and crystalline water ice at $\sim$40K. Again, Titan tholins were used
only to reproduce the spectral shape in the optical range (as in all
models of this work): this compound is certainly not present as such
at the surface of Orcus. Similarly, amorphous black carbon was only
used as a mean to match the low measured albedo. The corresponding
best-fit model is plotted in the bottom of Fig. \ref{fig:models}, the
quantities are tabulated in Table \ref{table:models} (model 1).

The large path lengths of amorphous water ice should be taken with
great care. Indeed, when alternatively using a blue component to fit
the continuum (using for instance the reflectance properties of
kaolinite from the ASTER database,
http://speclib.jpl.nasa.gov/search-1/mineral), the path lengths
\textit{and} the proportions of amorphous water ice in the final model
drastically drop. This indicates that crystalline water ice is most
probably dominating the mixture. All previous work on Orcus reports
the presence of water ice \citep{Fornasier+04,deBergh+05,Trujillo+05},
and the most recent higher SNR data show the presence of crystalline
water ice \citep[][and this work]{Barucci+08}. The crystalline water
ice fraction for this model is comparable to the one reported in
\cite{Barucci+08} (about 15$\%$). The depth of the absorption feature
at 1.65 $\mu$m depends on the crystalline water ice abundance and
temperature of the surface. At low temperature (close to that of
Orcus, $\sim$40K), a deep absorption feature as observed in our
spectrum (or in previous works) suggests a high
crystalline-to-amorphous water ice ratio.

This first attempt shows that a simple mixture of water ice
approximately reproduces the general spectral behavior of Orcus in the
near-infrared, but fails to describe the more subtle features (at
1.65$\mu$m and around 2.2$\mu$m). Orcus therefore has a water
dominated surface, but most probably hosts additional absorbing
substances, such as volatiles, in smaller amounts.

\subsection{Adding methane and ammonia ices}

For this second step, we tested various ices that could absorb around
2.2$\mu$m, contribute to the blue slope long-ward 2.3$\mu$m and could
possibly be present on a $\sim$40K KBO (see also
Fig. \ref{fig:components} and Sect. \ref{nir}). In particular, ammonia
hydrate has an absorption feature around 2.21$\mu$m: it was detected
on Uranus' satellite Miranda \citep{Bauer+02} and possibly identified
on Quaoar \citep{Jewitt+04} and Orcus \citep[previous work
by][]{Barucci+08}. Methane (dominating the surface composition of
large KBOs, such as Pluto, Eris or Makemake) is also an interesting
candidate, with its absorption feature around 2.2$\mu$m and blue
reflectance in the 2.3--2.4$\mu$m range. We also included methanol in
our tests (pure CH$_3$OH from Quirico \& Schmitt, priv. comm.)

At this stage, best-fit results were obtained using a mixture of water
ice (amorphous and crystalline), methane \textit{and} ammonia (pure or
diluted). The resulting models 2 and 3 are plotted in
Fig. \ref{fig:models} (see also Table \ref{table:models}). They have
an equivalent reduced $\chi^2$ and reproduce almost equally the
general shape of the near infrared spectrum.  The agreement of the
synthetic spectra to our data is however not totally satisfactory
around 1.65$\mu$m and 2.2$\mu$m, although the slope beyond 2.3$\mu$m
is now correctly mimicked. Only traces of methane and ammonia are
sufficient to significantly improve the fit with respect to a simple
water ice mixture. We tested the mixtures used by \cite{Barucci+08}
against our data, \textit{i.e.} trying separately methane, then
ammonia hydrate (respectively mixed with water ice) but obtained a
worse fit than models 2 and 3. In particular, the 2.2$\mu$m band was
poorly reproduced in both cases, and for the ammonia hydrate model,
the synthetic spectrum had a reflectance that was too high beyond
2.3$\mu$m.

From a theoretical point of view, ammonia and methane could be major
volatiles in the solar nebula at the distance where Orcus might have
formed \citep{Lewis72}. Ammonia on one hand is known to lower the
temperature of the melting point when mixed with water ice. Methane on
the other hand is almost insoluble in aqueous liquids at low pressures
(no electric dipole moment). At the central pressure of Orcus (few
kilo-bars), its solubility should nonetheless reach about 1\%. Methane
could therefore trigger so-called explosive aqueous cryovolcanism in
the presence of ammonia and water \citep{Kargel+92, Kargel+95}. This
process has been introduced to explain the cryovolcanism observed on
Triton \citep{Kargel+90}. Indeed, the bubble content of liquid melts
would increase as they propagate toward the surface through cracks:
the solubility of methane in these melts depend on the pressure, which
decreases as the melts reach the surface. If the bubble content
becomes sufficient, the melts would eventually disintegrate into a
gas-driven spray producing the explosive cryovolcanism.

\subsection{Search for daughter species of methane and ammonia}

Ethane is an irradiation product (from UV photolysis) of solid ethane
\citep{Baratta+03}, and is expected to be involatile at
$\sim$40K. Therefore, once produced by solar irradiation, it should
remain at the surface of the object and still be
observable. \cite{Moore+03} report that the production of C$_2$H$_6$
is more efficient when CH$_4$ is pure. C$_2$H$_6$ was suggested at the
surface of Quaoar \citep{SchallerQuaoar+07} and detected on dwarf
planet Makemake \citep{Brown+07}. Since our data could be compatible
with traces of methane at the surface of Orcus, we decided to further
investigate the contribution of C$_2$H$_6$. We used the optical
constants from \cite{Quirico+97}. Following a similar reasoning, we
decided to test the presence of ammonium (NH$_4^+$), which can be
produced for instance by irradiation of pure NH$_3$ or mixtures of
water and ammonia as experienced by \citep{Moore+07}. Other species
can be produced from the irradiation of NH$_3$ or NH$_3$:H$_2$O, such
as hydroxylamine (NH$_2$OH); unfortunately, the corresponding weak
infrared features are not easily detectable. In this test, we used
optical constants of NH$_4^+$ derived by J. Cook (priv. comm.) from
the studies by Moore et al.

We therefore tested ammonia and methane separately along with their
respective irradiation products. Model 4 (middle quadrant of Fig.
\ref{fig:models}) shows a mixture of water ice, methane, and ethane,
while model 5 includes water ice, ammonia hydrate, and ammonium. The
methane option (model 4) gives a lower reduced $\chi2$ and overall
better fit than model 5, and is visually and statistically equivalent
to models 2 and 3. However, ethane is finally found not to contribute
to the final mixture (0\%) when mixed with only CH$_4$ and water ice,
which tells us that methane is itself an interesting component to
describe the H and K band data. The ammonia and ammonium mixture model
is not satisfactory to improve the overall fit of Orcus spectrum (with
respect to other models), but the structure of the 2.2$\mu$m band is
worth of interest.

\subsection{Best-fit results and conclusions}

For this last step we aimed at better describing the data overall, in
particular in the 2.2$\mu$m region, using results (and possible
candidates) derived from the previous investigation steps. Our
best-fit result overall is displayed as model 6 (top of
Fig. \ref{fig:models}) and includes a mixture of water ice mostly in
its crystalline state ($\sim$20\%), ammonium ($\sim$13\%) and traces
of ethane ($\sim$7\%). The overall shape of the Orcus spectrum is
correctly reproduced, in particular in the K band with both the
2.05$\mu$m large water ice band and the 2.2$\mu$m region more
correctly fitted. The presence of ammonium also better reproduces the
visible slope. The 1.65$\mu$m feature was never completely correctly
reproduced, which could be related to the SNR of the data.  Our
conclusion is that the irradiation products NH$_4^+$ and C$_2$H$_6$
greatly helps to describe the data in the 1.4--2.4$\mu$m region,
although both parent species, as a mixture of ammonia (pure or
diluted) and methane cannot be firmly excluded (model 2 and 3 of
equivalent statistical level). We believe that the four species are
probably present as traces at the surface of Orcus. Therefore, further
spectroscopic studies of Orcus with significanlty higher SNR
(especially in the 2.2$\mu$m region) would improve our knowledge about
the presence of the various ices (such as methane and ammonia) and
related chemical products and processes on the surface of this object.

\subsection{Surface composition variations over time?}

We saw that the neutral, featureless (within the noise level) visible
spectral properties are consistent over a timespan of about 4
years. It is almost impossible to phase the different spectra with the
rotational ligthcurves available because they were not observed close
enough in time. However, as Orcus does not show dramatic lightcurve
effects (see Sect. \ref{sec:intro}) and is expected to be a spherical
object, it is reasonable to think that the surface properties are
rather homogeneous over the surface in the optical. In the
near-infrared, according to the most recent, higher SNR data
\citep[][and this work]{Barucci+08}, crystalline water ice was always
present (and in the same roughly proportions as in the indicative
Hapke modeling). A 2.2$\mu$m feature was also systematically shown to
be there in the most recent works, with slightly different structures,
although we believe they are roughly consistent within the
noise. Until significantly higher SNR data are available and show more
subtle changes (as for instance Charon), we can conclude that Orcus'
spectral properties are rather homogeneous over its surface in the
near-infrared at the currently available resolution.

\subsection{Contribution of the satellite}

The known properties of the Orcus binary system are summarized in
Table \ref{tab:vanth}. Because of the small separation of the pair
\citep[0.26$\arcsec$][]{Noll+08}, it is impossible to separate the
contribution of both components in our ground-based observations.  The
color measurements performed with the \textit{Hubble Space Telescope}
(from which the binary system can be resolved) by \cite{Brown+10} show
that Vanth displays significantly redder visible \textit{and}
near-infrared colors than Orcus. The contribution of the satellite to
the total flux is 8\% in the visible and a bit more in the
near-infrared. Thus, we do not expect large variations in the water
ice content of the main body. Ammonia, methane, and their irradiation
products, if confirmed, should be present at the surface of the
primary object only, as Vanth is probably too small to have retained
volatile ices at its surface. Given that the satellite's near-infrared
color index is redder than that of Orcus \citep[as measured
by][]{Brown+10}, we can expect that the satellite will have a lower
water ice content, if any.

\section{Comparison with other water-bearing KBOs of interest}

%_____________________________________________________________
\begin{table*}[!th]
\begin{minipage}[t]{2.\columnwidth}
% \begin{minipage}[t]{\columnwidth}
 \caption{Comparison of orbital and physical properties of some
    intermediate-large size water-bearing objects of
    interest. References are: (1) \cite{Stansberry+08}, (2)
    \cite{Sicardy+06} and references therein, (3) \cite{Brown+10}, (4)
    \cite{Brown+04}, (5) \cite{Brown08}, (6) \cite{Brucker+09}, (7)
    computed from \cite{Buie+10}, (8) \cite{Rabinowitz+06}, (9)
    \cite{Cook+09}, (10) \cite{Guilbert+09}, (11)
    \cite{SchallerQuaoar+07}, (12) \cite{Dalle+09}, (13)
    \cite{MerlinCharon+10} and references therein.}
\label{tab:objcompare}      
\centering           
\renewcommand{\footnoterule}{}  % to avoid a line before footnotes
\begin{tabular}{l l l l l l} 
\hline          
\hline   
Parameter                                                           &Orcus            & Charon           & (208996) 2003 AZ$_{84}$        &Quaoar           &Haumea\\   
\hline
Dynamical Class\footnote{Using classification by \cite{Gladman+08}} &3:2              &3:2               & 3:2           & Classical       & Classical\\
Orbit inclination ($^{\circ}$)\footnote{with respect to ecliptic plane, from the Minor Planet Center}        &20.6             &17.1              &13.6           &8.0              &28.2\\
Semi-major axis (AU)\footnote{From the Minor Planet Center}         &39.2             &39.6              &39.4           &43.5             &43.0\\ 
q (AU)\footnote{Perihelion from the Minor Planet Center}            &30.2             &29.7              &32.3           &41.7             &34.5\\ 
Q (AU)\footnote{Aphelion from the Minor Planet Center}              &48.1             &49.6              &46.5           &45.2             &51.5\\
r (AU)\footnote{Approximate current heliocentric distance}          &47               &31                &45             &43               &51\\
\hline
Diameter (km)                                                       &946$\pm70^{(1)}$   &1207$\pm3^{(2)}$    &686$\pm95^{(1)}$  &850$\pm200^{(1)}$  &1150$\pm200^{(1)}$\\
                                                                    &940$\pm70^{(3)}$   &                  &                &910$\pm100^{(6)}$  &$\sim 2000\times1500\times1000^{(5)}$\\
                                                                    &                &                   &               &1260$\pm190^{(4)}$ &\\
\hline
p$_V$ (\%) \footnote{Visible geometric albedo}                      &19.7$\pm3^{(1)}$   &$\sim40^{(7)}$             &12.3$\pm4^{(1)}$  &19.9$\pm10^{(1)}$  &80$\pm15^{(1)}$ \\
                                                                    &28$\pm4^{(3)}$     &                   &                &17.2$\pm5^{(6)}$        &60-80$^{(8)}$\\
                                                                    &                 &                   &                &9.2$\pm4^{(4)}$ &\\
\hline
Vis. slope (\%/100nm)\footnote{Slope of the visible spectrum from \cite{Fornasier+09} and references therein, except for Orcus: this paper}       &$\sim2$             &---      &$\sim3$ &$\sim28$&$\sim0$\\
Dominant ice at surface                                             &H$_2$O (cryst)       &H$_2$O (cryst)                       &H$_2$O (cryst)  &H$_2$O (cryst)             &H$_2$O (cryst)\\
Volatiles and other                                                 &NH$_4^+$+C$_2$H$_6$  &NH$_4^+$+CH$_4$+NH$_3$:H$_2$O$^{(9)}$     &CH$_3$OH ?$^{(10)}$  &CH$_4$+C$_2$H$_6^{(11)}$       &---\\
                                                                    &NH$_3$:H$_2$O+CH$_4$ &NH$_3$:H$_2$O, NH$_3$$^{(13)}$           &           &CH$_4$+C$_2$H$_6$+N$_2^{(12)}$ &\\
                                                                    &NH$_3$+CH$_4$        & NH$_4$OH$^{(13)}$&&&\\
\hline
\end{tabular}
\end{minipage}
\end{table*}
%_____________________________________________________________

In this section we look at the surface expression of water (more
specifically in its crystalline form) and additional traces of
volatiles on some other large-intermediate water bearing objects. We
try to point out common properties and differences, and in the case of
Quaoar, infer general clues about its possible history and
interior. The compared properties are summarized in Table
\ref{tab:objcompare}.

\subsection{(50000) Quaoar}

The KBOs (90482) Orcus and (50000) Quaoar seem to share the same type
of near-infrared spectrum: dominated by crystalline water ice, with
small amounts of volatiles \citep{SchallerQuaoar+07}. Besides, they
are both intermediate-size KBOs with similar albedo
\citep{Stansberry+08}: their study could thus shed light on the
possible transition between large methane-dominated objects (like
Pluto, Eris, Sedna and Makemake) and smaller water-ice-dominated
ones. Nonetheless, these two objects might be very different when it
comes to their thermal and dynamical history. Moreover, their visible
spectral properties differ: Quaoar, with its $\sim$28\%/100nm slope,
is part of the objects hosting ultra-red material
\citep[$>$25\%/100nm,][]{Jewitt02} on its surface. Table
\ref{tab:objcompare} and \ref{tableQ} show a comparison between
various physical and dynamical properties of Orcus and Quaoar.
\\

In this table, the maximum surface temperature is computed from a
thermal balance for the whole surface including different energy
inputs such as: the incoming solar energy,
\begin{equation}
E_{\odot} = (1-\mathcal{A})\frac{C_{\odot}}{d_H^2}cos(\xi), 
\end{equation}

(with $\mathcal{A}$ the Bond albedo, $C_{\odot}$ the solar constant,
$d_H$ the heliocentric distance and $\xi$ the local zenith angle), the
thermal emission $\varepsilon \sigma T^4$ (with $\varepsilon$ the
thermal emissivity, $\sigma$ the Stefan-Boltzmann constant and $T$ the
temperature), the radial heat flux and lateral heat fluxes,
considering a thermal conductivity of about
10$^{-2}$~W.m$^{-1}$.K$^{-1}$. This thermal balance is computed for
about 50 orbits with a full 3D thermal evolution model to get the
resulting temperatures \citep[see Appendix B of][]{Guilbert+10}.  The
insulation per orbit (in J), is computed with the formula
\citep{Prialnik+04}
\begin{equation}
  \mathcal{Q}= \frac{\pi L_{\odot}}{2\sqrt{GM_{\odot}}} \frac{R^2}{\sqrt{a(1-e^2)}},
\end{equation}
with $L_{\odot}$ and $M_{\odot}$ the solar luminosity and mass
respectively, $G$ the gravitational constant and $R$ the object's
radius. Depending on the radius value we use for Quaoar, two different
values are obtained for Q. The values for Orcus and Quaoar (and
surface temperature) are consistent, although their spectral response
to solar illumination is very different. Indeed, despite the similar
visual albedo and near infrared spectral properties, Orcus' visible
spectrum is neutral, whereas Quaoar's spectrum has a red slope: they
most probably have different thermal properties, in terms of their
spectral response to solar illumination, light absorption and
subsequent heating.
\\

Quaoar is a classical KBO and as such should have been formed
\textit{in situ}, or was moderately pushed outward during Neptune's
migration \citep{Gomes+03}. Therefore, it should have been formed on a
larger timescale than Orcus, implying potentially less accretional
heating. Orcus is a Plutino and might have been formed closer to the
Sun than its current location \citep{Malhotra93, Malhotra95}. We can
thus suspect that its accretion was faster and involved more heat. The
early thermal history of both objects in this sense might be quite
different. Nonetheless, because of their rather large sizes and
densities, they could have both reached temperatures sufficient to
melt water ice and produce a differentiated internal structure
(\textit{i. e.} after heating by short-lived radiogenic elements, the
interior should be stratified, with a differentiated core and layers
enriched/impoverished in volatile ices closer from the surface).
\\

The abundance of radiogenic elements is directly related to the dust
mass fraction within the body: the denser an icy body is, the more
heat will be released upon the decay of these radiogenic
elements. Considering Quaoar's high density, we can assess that it
might have suffered from severe thermal modifications due to the decay
of long-lived radiogenic elements such as ${}^{40}$K, ${}^{232}$Th,
${}^{235}$U, ${}^{238}$U. Therefore, the methane observed at its
surface could have been placed there in a quite recent past, although
the object should be geologically dead for millions years. The
subsequent surface evolution should have been dominated by
space-weathering processes. Orcus is less dense than Quaoar: a thermal
evolution model is applied to test if long-lived radiogenic elements
could have heated the object in the last million years, so that the
effects on the surface would still be observable today (see Sect.
\ref{thermalModel}).

%_____________________________________________________________
%
\begin{table}
  \caption{Comparison of Orcus and Quaoar additional parameters. 1: values from the Minor Planet Center. References are: a: \cite{Brown08}, b: \cite{Brown+10}, c: \cite{Fraser+09}}             
\label{tableQ}     
\centering          
\begin{tabular}{c|lll}  
  \hline 
\hline    
  & \textbf{Parameter} & \textbf{Orcus} & \textbf{Quaoar}\\ 
  \hline 
  \multirow{4}*{\rotatebox{90}{Orbit}}
  & Dynamical class                     & Plutino                             & Classical KBO                \\ 
  & Semi-major axis$^{1}$ (AU)          & 39.163                              & 43.467                       \\
  & Eccentricity$^{1}$                  & 0.227                               & 0.040                         \\
  & $P$ (yrs)                           & 245.5                               & 286.9                         \\ 
\hline
  \multirow{5}*{\rotatebox{90}{Phys. prop.}}
  & \multirow{2}*{Density (g.cm$^{-1}$)} & 1.9$\pm$0.4$^{a}$                    & \multirow{2}*{$>$2.8$^{c}$}        \\
  &                                    & 1.5$\pm$0.3$^{b}$                    &                                               \\
\cline{2-4}
  & T$_{surf}$ max. (K) & 40-50 & 42-46 \\
\cline{2-4} 
    & \multirow{2}*{Insulation per orb. (J)}   & \multirow{2}*{$\sim$2.0$\times$10$^{21}$}   & 1.5$\times$10$^{21}$ \\
  &                                    &                                       & 3.3$\times$10$^{21}$ \\ 
  \hline 
  \multirow{2}*{\rotatebox{90}{Sat.}}
  & Brightness                         & \multirow{2}*{8$^{a}$}              & \multirow{2}*{0.6$^{a}$}\\
  &fraction (\%)&&\\
  \hline
\end{tabular}
\end{table}
%
%_____________________________________________________________

\subsection{The Haumea collisional family}

(136108) Haumea is a 2000 km-class fast-rotating elongated dwarf
planet from the classical Kuiper Belt with the highest water ice
surface content currently known among KBOs and a neutral visible
spectrum \citep[see][and references
therein]{Pinilla+09,Merlin+07}. \cite{BrownNat+07} identified several
objects with similar spectral properties in a dynamical area that is
close to Haumea. Because Haumea's unusual rotational properties are
compatible with a giant impact, it was soon suggested that these
objects are members of the first collisional family identified in the
Kuiper Belt. Repeated spectral and photometric studies
\citep{Snodgrass+10,Barkume+08,Rabinowitz+08,Pinilla+08,Schaller+07,Pinilla+07}
aimed at refining the family properties and members list.

From these studies, we see that despite the presence of water ice and
the neutral visible spectrum, the other properties of the Haumea
collisional family differ from those of Orcus. Haumea itself has a
much higher albedo \citep[0.6--0.84][]{Rabinowitz+06,Stansberry+08},
and density \citep[2600--3340 kg m$^{-3}$][]{Rabinowitz+06} and much
deeper water ice absorption bands.  \cite{Pinilla+09} showed that it
has a spectrally homogeneous surface best described by a simple
intimate mixture of amorphous and crystalline water ice and that it is
probably depleted of any other ices. As a consequence, cryovolcanism
cannot be invoked as possible a re-surfacing process, whereas it might
play a role in Orcus' evolution. In this work, we also showed that a
simple water ice mixture (Haumea-type surface) fails to completely
reproduce Orcus' spectral behavior in the near-infrared (see model 1).

\subsection{Charon}

Pluto's larger satellite Charon is a $\sim$1200 km size body with a
water ice dominated surface. Charon and Orcus are of similar size and
density (1.5--1.9 g.cm$^{-1}$), on a similar orbit, and share common
spectral properties. However, their albedos are quite different, as
Charon was measured to have a visible geometric albedo of $\sim$0.40
\citep[computed from][]{Buie+10}. Crystalline water ice is
unambiguously detected and leads to a surface temperature estimate of
40--50K \citep{Cook+07}, which is in the same range as that of Orcus
\citep[44K from][]{Barucci+08}. Another component absorbing at
2.21$\mu$m is preferably attributed to ammonia-type ices from detailed
Hapke modeling \citep[][and references therein]{MerlinCharon+10}, and
it seems that the surface is variegated (differences between the Pluto
and anti-Pluto hemispheres). Recent high SNR K-band spectral studies
\citep{Cook+09} suggested for the first time that ammonium could be
present at the surface of Charon, as it helps to better model the
absorption features they detected around 2.21 and 2.24$\mu$m. Their
best-fit Hapke model includes a mixture of water, ammonium, methane,
and ammonia hydrate. In this work, we tried to fit to our data a
similar synthetic spectrum: we obtained a reduced $\chi^2$ of 1.68
(\textit{e.g } of lower quality than our best-fit options), although
the general spectral features of Orcus were correctly reproduced. This
confirms the similarities of Orcus and Charon bulk surface
properties. Also, \cite{Desch+09} identified from their 1D-spherical
thermal model of Charon that cryovolcanism is the most probable
mechanism to provide fresh ices on the surface. From our own thermal
model (see Sect. \ref{thermalModel}), we draw a similar conclusion
for the case of Orcus.

\subsection{Plutino (208996) 2003~AZ$_{84}$ }

(208996) 2003~AZ$_{84}$ is, like Orcus, a binary Plutino wandering in a
30--50 AU heliocentric region. It is presently located at a similar
heliocentric distance, $\sim$46 AU (wrt $\sim$48 AU for
Orcus). (208996) has a quasi-neutral featureless spectrum in the
optical, with a spectral slope of 1.6~$\pm$~0.6\%/100nm \citep[][and
references therein]{Fornasier+09}. \cite{Guilbert+09} confirmed the
presence of water ice \citep{Barkume+08} in its crystalline phase from
near-infrared spectroscopy, and noted a possible absorption feature
around 2.3$\mu$m. The SNR of the data is not sufficient to identify
the corresponding absorbing substance. However, the general spectral
behavior of Plutino (208996) is close to that of Orcus, and joint
studies deserve to be conducted in the future.

\section{A thermal-evolution model of Orcus}
\label{thermalModel}
 
\subsection{The model}
We used a thermal-evolution model in order to evaluate the possibility
for Orcus to undergo some internal activity events such as
cryovolcanism in a recent past, so that the effects might still be
observable at the surface. From input orbit, size, albedo, density,
porosity, formation delay and type of radiogenic elements, we
simulated the evolution of temperature over the age of the Solar
System for the whole object. In this model, the heat diffusion
equation (Eq. \ref{eq_depart}) is solved using a fully 3D scheme,
considering internal heating due to the crystallization of amorphous
water ice, as well as heating due to the decay of several short- and
long-lived radiogenic elements. We therefore need to solve the
equation
\begin{equation}
\label{eq_depart}
\rho _{bulk}c~\frac{\partial T}{\partial t}~+~div(- \kappa~ \overrightarrow{grad}~T)~ =~\mathcal{S},
\end{equation}
where $T$ is the temperature distribution we need to determine, $\rho
_{bulk}$ the object's bulk density, $c$ the material heat capacity,
$\kappa$ its effective thermal conductivity and
$\mathcal{S}=\mathcal{Q}_{rad}+\mathcal{Q}_{cryst}$ the two heat
sources. The radiogenic heat source is described by
\begin{equation}
\mathcal{Q}_{rad} = \sum _{rad}\varrho _{d} ~ X_{rad}~ H_{rad}\frac{1}{ \tau _{rad}} \exp \left( \frac{-t}{\tau _{rad}}\right), 
\end{equation}
with $\varrho _{d}$ the dust bulk density, $X_{rad}$ the initial mass
fraction of a particular radiogenic element in the dust, $H_{rad}$ the
heat released by this radiogenic element per unit mass upon decay,
$\tau _{rad}$ its decay time, and $t$ the time. We consider different
short- and long-lived radiogenic elements, presented in Table
\ref{elements}, using data adapted from \citet[][and references
therein]{Castillo+07} for ordinary chondrites.

%--------------------
\begin{table}
\begin{minipage}[t]{\columnwidth}
\caption{Physical parameters of various short-period
  (SP) and long-period (LP) radiogenic elements considered in the
  model. Values adapted from \citet{Castillo+07}.}
\label{elements}
\centering
\renewcommand{\footnoterule}{}
\begin{tabular}{ccccc}
\hline
\hline
   &  \multirow{2}*{Element} & \multirow{2}*{$X_{rad}$(t=0)\footnote{Initial mass fraction of the radiogenic element}} & $\tau _{rad}$\footnote{Decay time} & $H_{rad}$\footnote{Heat released by this radiogenic element per unit mass upon decay} \\
   &                  &                       &   (years)     &        J.kg$^{-1}$           \\
\hline 
\multirow{3}*{SP}
&~{}$^{26}$Al    & 6.7$\times$10$^{-7}$    &1.05$\times$10$^{6}$&4.84$\times$10$^{12}$\\
&~{}$^{60}$Fe   & ~~~~2.5$\times$10$^{-7/-8}$&2.16$\times$10$^{6}$&5.04$\times$10$^{12}$\\
&~~{}$^{53}$Mn  & 2.8$\times$10$^{-8}$    &5.34$\times$10$^{6}$&4.55$\times$10$^{12}$\\
\hline 
\multirow{4}*{LP}
&{}$^{40}$K    & 1.2$\times$10$^{-6}$&1.80$\times$10$^{9}$  &1.66$\times$10$^{12}$\\
&{}$^{232}$Th& 6.0$\times$10$^{-8}$&~2.02$\times$10$^{10}$&1.68$\times$10$^{13}$\\
&{}$^{235}$U~  & 9.0$\times$10$^{-9}$&1.02$\times$10$^{9}$  &1.82$\times$10$^{13}$\\
&{}$^{238}$U~  & 2.9$\times$10$^{-8}$&6.45$\times$10$^{9}$  &1.92$\times$10$^{13}$\\
\hline
\end{tabular}
\end{minipage}
\end{table}
%---------------

The heat source due to the crystallization of amorphous water ice is
described by
\begin{equation}
\mathcal{Q}_{cryst} = \lambda (T)~ \varrho _{a}~ H_{ac},
\end{equation}
with $\varrho _a$ the amorphous water ice bulk density. This phase
transition releases a latent heat $H_{ac}$, whose value is
9$\times$10$^{4}$~J.kg$^{-1}$ \citep{Klinger81}, with a rate measured by
\citet{Schmitt+89} of:
\begin{equation}
\lambda (T) = 1.05 \times 10^{13} ~ e^{-5370/T} ~~s^{-1}.
\end{equation} 

The heat diffusion equation is then expanded on the real spherical
harmonics basis. The resulting equation is numerically solved using an
implicit stable Crank Nicolson scheme. A full description of this
model can be found in \cite{Guilbert+10}.

Several physical parameters have to be initialized before simulating
the thermal evolution of Orcus. We first assume that the object is on
its current orbit for the whole simulation and is placed there right
after its formation time. This does not reflect the true dynamical
evolution of Orcus, but might not strongly affect the resulting
general thermal behavior (as detailed in the next paragraphs). Orcus
is then assumed to have a 475~km radius, a 19.9\% albedo
\citep{Stansberry+08}, and a bulk density of 1900~kg.m$^{-3}$
\citep{Brown08}. By assuming a residual porosity of 10\%
\citep{Durham+05}, this density leads to a mixture made of 77\% of
dust and 23\% of water ice (mass fractions), the dust being
homogeneously distributed within the ice matrix. The resulting thermal
properties are: the thermal conductivity
$\kappa~=~0.18~$Wm$^{-1}$K$^{-1}$, the heat capacity
$c~=~10^3$~Jkg$^{-1}$K$^{-1}$.

Since heating due to the accumulation of gravitational potential
energy is usually not important, even for large KBOs such as Pluto
\citep{McKinnon+97}, we neglect this heat source in this model. We
also assume that the objects undergo a cold accretion: the accretional
heating is thus neglected, although it could be a very important heat
source in the case of Orcus.

\subsection{Results}
\label{sec:modelresults}

\subsubsection{Effect of short-lived radiogenic elements}

The formation time of KBOs is of critical importance when it comes to
the study of their differentiation due to heating by short-lived
radiogenic elements. These elements indeed decay during the accretion
period, thus providing much less potential heat when the bodies are
formed (and our simulations start). There is still a wide range of
available values for this formation time, depending on models. For
example, \citet{Wei+04} suggested that 50~km-radius objects could be
formed in less than one million years inside 30~AU. Recent simulations
performed by \citet{Kenyon+08} show that 1000~km-radius bodies could
be formed in 5-10$\times$10$^6$~yrs within the 20-25~AU region.
\\

With the aim of constraining Orcus' state of differentiation (primordial
and during its lifetime), we applied the model considering various
formation delays as input. The initial temperature is 30~K (cold
accretion), and the radiogenic elements initial mass fraction is
calculated using the following formula, to account for their decay
during accretion
\begin{equation}
X_{rad}(t_F)=X_{rad}(0)~e^{-t_F/\tau_{rad}},
\end{equation}
with $t_F$ the considered formation delay, and $X_{rad}(0)$ found in
Table \ref{elements}. All SP and LP elements listed there are taken as
input, and we distinguish the case of the high and low $^{60}$Fe mass
fraction. Our simulations over the age of the Solar System
(4.5$\times$10$^9$~yrs) show that Orcus could have a physically
differentiated structure --including a melted core, crystalline water
ice mantle, and a pristine surface layer composed of amorphous ice and
dust-- if formed in less than about 5$\times$10$^6$~yrs. The melting
point of water is indeed achieved within the body in the first
millions years of the simulations. If we assume that ammonia is also
present in the mixture, then this formation time limit is slightly
increased to about 7$\times$10$^6$~years, since the presence of
ammonia depresses the melting point from a few K to almost 100~K
depending on its abundance \citep[][and references
therein]{Kargel+92}.
\\
% - - - - - - - - - - - - - - - - - - - - - - - - - - - - - - 
\begin{figure}
\begin{center}
   \resizebox{\hsize}{!}{\includegraphics{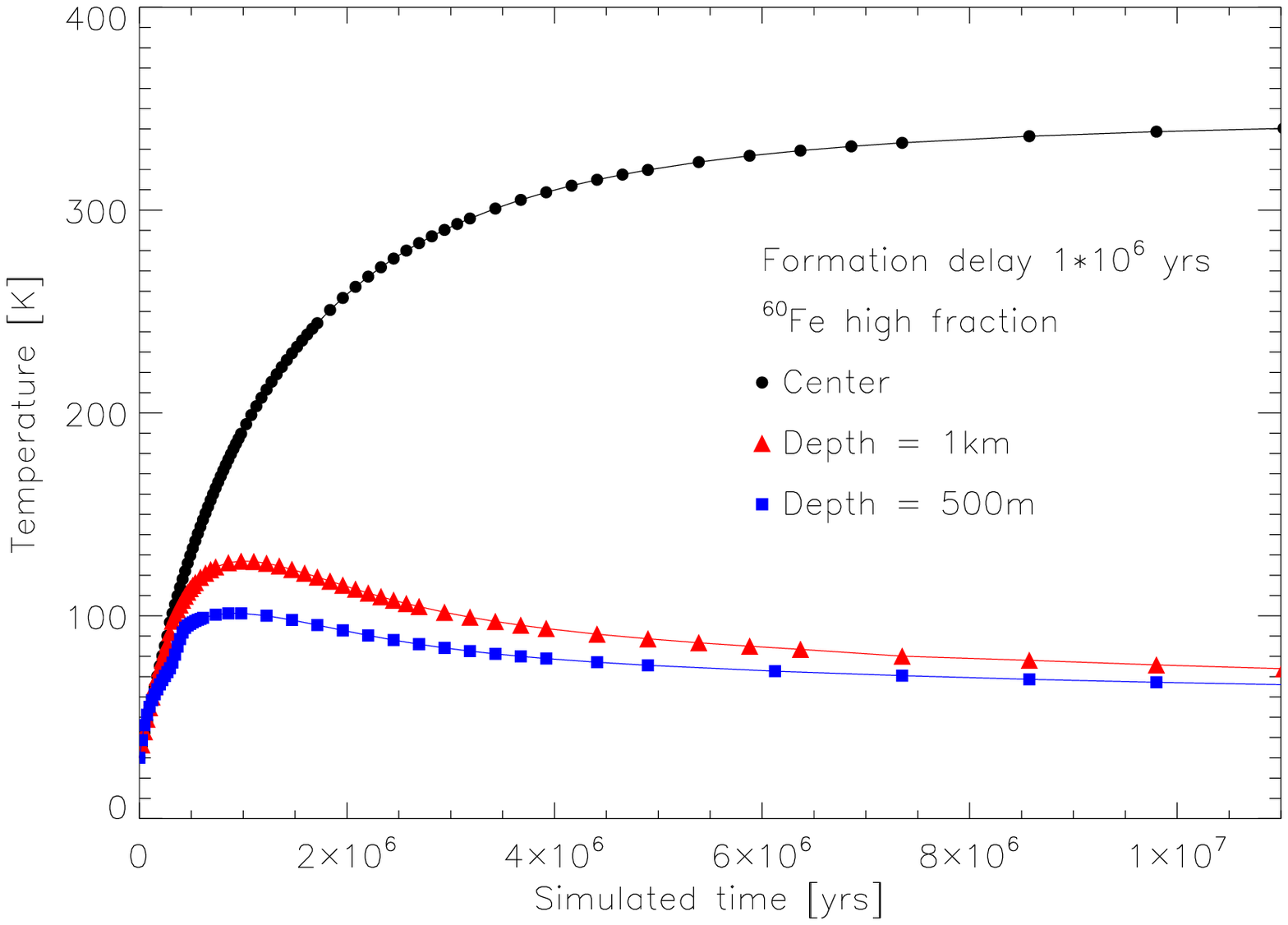}}
   \resizebox{\hsize}{!}{\includegraphics{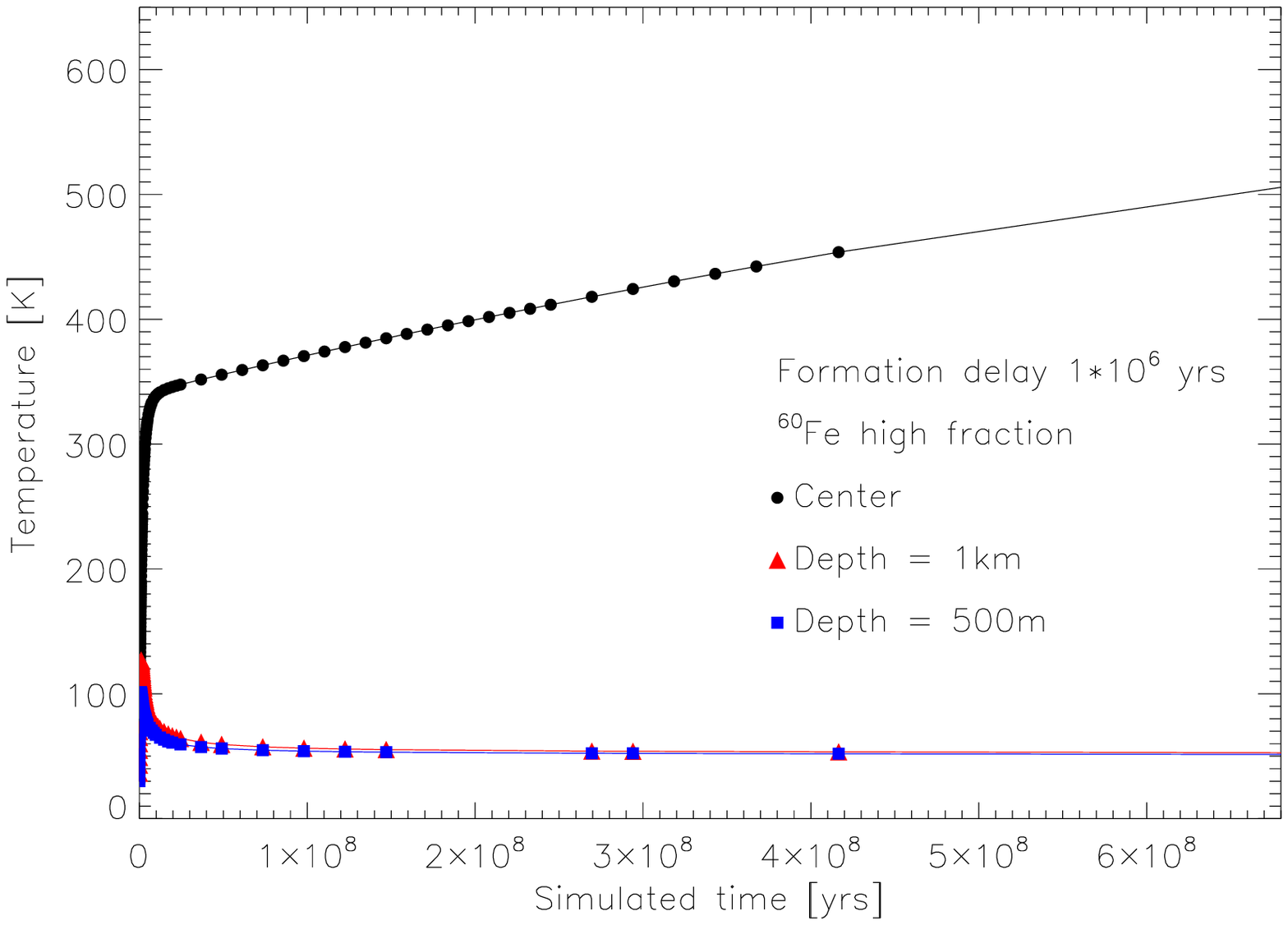}}
   \caption{Evolution of Orcus' central temperature (black circles) as
     a function of the simulated time in the case of a formation delay
     of 1 million years and a high mass fraction of radiogenic element
     ${}^{60}$Fe. Triangles and square curves correspond to the
     temperature at a depth of 1km and 500m resp. Top graph zooms
     on the first $\sim$10 Myr of simulation.}
\label{fig:orcus-1M}
  \end{center}
\end{figure}   
% - - - - - - - - - - - - - - - - - - - - - - - - - - - - - - 

Figure \ref{fig:orcus-1M} shows the temperature evolution inside Orcus
at various depths (center, 1km-deep, 500m-deep) considering the
highest mass fraction of short-lived ${}^{60}$Fe and a formation delay
of 1 Myr. Although this formation delay might seem too short for such
a large object, the corresponding results show the greatest
temperature variations and the steepest thermal gradients expected, to
put an upper limit to all thermal processes. The radiogenic heating is
homogeneous inside the body, but the surface is in thermal equilibrium
with the external medium. This results in the propagation of a cold
wave from the surface toward deeper layers. As a consequence, the
curves corresponding to the 1km-depth and 500m-depth show a maximum
temperature after a few millions years of the simulations: the body
then starts to cool slowly.
\\
% - - - - - - - - - - - - - - - - - - - - - - - - - - - - - - 
\begin{figure}
\begin{center}
   \resizebox{\hsize}{!}{\includegraphics{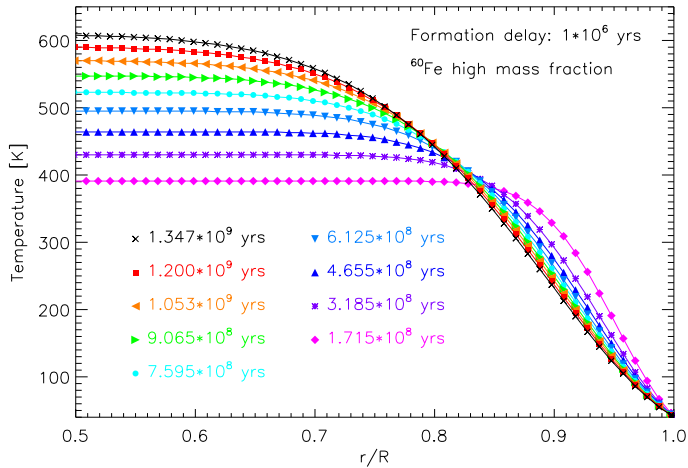}}
   \caption{Temperature (K) as a function of the object's depth for
     various evolution times. This plot illustrates how the outside
     medium cools the surface layers deeper with time (for SP and LP
     radiogenic elements, high $^{60}$Fe mass fraction.)}
\label{fig:coupeSP}
  \end{center}
\end{figure}   
% - - - - - - - - - - - - - - - - - - - - - - - - - - - - - - 

Another consequence is that a layer of amorphous water ice is
maintained at the surface of the object: the 1km-deep layer
temperature reaches the water crystallization threshold at some point,
while the 500m-deep layer does not complete the crystallization
process. The propagation of the cold wave from the surface to the
interior is illustrated by Fig. \ref{fig:coupeSP}, which shows the
radial temperature profile across the body for different simulated
times. The thickness of this pristine material layer depends on the
formation delay that we consider: the shorter the delay, the thinner
the layer. We find that the thickness could be as low as several
hundreds of meters ($\sim$500m for the 1 Myr delay) or as thick as a
few tens of kilometers ($\sim$10km for the 10 Myr delay). Most of the
volume is therefore made of crystalline water ice. Our simulations
also show that the threshold for the solid-liquid phase transition of
water is reached within most of the volume (if formed fast enough):
the high temperatures needed to keep liquid water inside Orcus are
moreover maintained for long periods such as 2$\times$10$^9$~years.

\subsubsection{Effect of long-lived radiogenic elements}  

% - - - - - - - - - - - - - - - - - - - - - - - - - - - - - - 
\begin{figure}
\begin{center}
   \resizebox{\hsize}{!}{\includegraphics{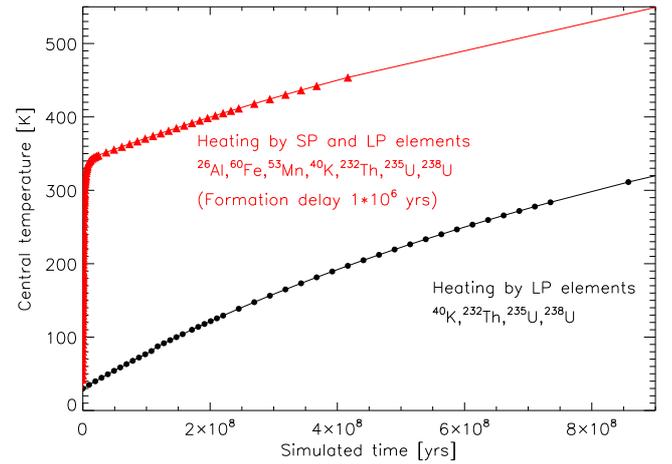}}
   \caption{Central temperature (K) as a function of the simulated
     time, showing the different contributions from SP+LP radiogenic
     elements, and LP only.}
\label{fig:SPLP}
  \end{center}
\end{figure}   
% - - - - - - - - - - - - - - - - - - - - - - - - - - - - - - 

What would happen if Orcus was formed in more than 10 Myr? In this
case, the effect of short-lived radiogenic elements would be
nonexistent. However, radiogenic heating due to the decay of
${}^{40}$K, ${}^{232}$Th, ${}^{235}$U, and ${}^{238}$U would still be
inevitable. This would even be of the greatest importance for Orcus,
whose thermal timescale,
\begin{equation}
\tau_{th} = \frac{R^2\rho_{bulk} c}{\pi^2\kappa}, 
\end{equation}
exceeds the age of the Solar System (assuming the previous thermal
physical parameters). Nonetheless, the value of the thermal
conductivity is critical for this calculation, and a variation of one
magnitude could reduce the cooling time to less than 10$^9$~years.

\begin{figure}
\begin{center}
   \resizebox{\hsize}{!}{\includegraphics{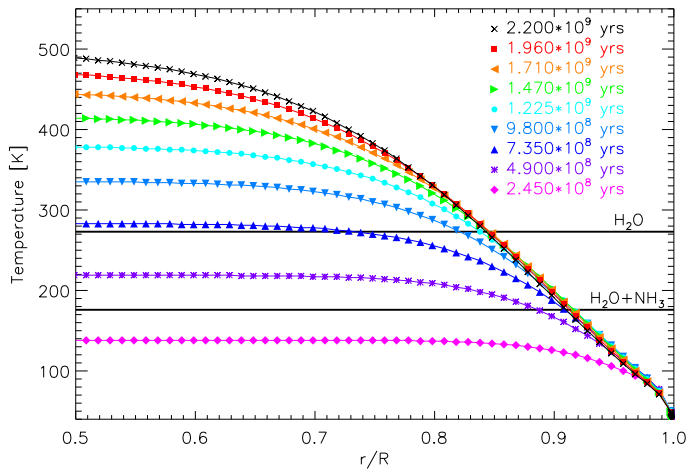}}
   \caption{Temperature (K) as a function of the object's depth for
     various evolution times (the lower limit is the case of
     radiogenic heat source from long-lived elements only). Horizontal
     lines show the temperature above which H$_2$O or H$_2$O+NH$_3$
     are in the liquid phase.}
\label{fig:coupe-z}
  \end{center}
\end{figure}

We therefore applied our model including only the long-lived
radiogenic elements ${}^{40}$K, ${}^{232}$Th, ${}^{235}$U, and
${}^{238}$U. The evolution of the body's central temperature is shown
on Fig. \ref{fig:SPLP}, compared to the evolution under the effect of
both long- and short-lived elements. Even without the high heating
power of short-lived elements, the internal temperature increases,
reaching the melting point of water ice within the first billion
years. Please keep in mind that our mixture contains a high fraction
of dust that might not reflect the true composition of Orcus. In this
case Orcus suffers from considerable heating, while it could actually
be much more moderate. Our simulations show that in the case of an
important heating by long-lived radiogenic elements, the melting point
of water ice could be reached within about 60\% of the body, while the
melting point of an ammonia-water mixture would be reached within 80\%
of the body. This is illustrated in Fig. \ref{fig:coupe-z}, which
shows radial temperature profiles for different simulated times. As in
the cases considered in the previous section, a layer of amorphous
water ice is maintained at the surface of Orcus. Its thickness is
about 5-6~km.

% - - - - - - - - - - - - - - - - - - - - - - - - - - - - - - 
% - - - - - - - - - - - - - - - - - - - - - - - - - - - - - - 

\subsection{Implications}

\subsubsection{Limitations}

Our model has several limitations that need to be kept in mind when
interpreting the results. The main issue is that the liquid phase or
differentiation is not accounted for: the model is not suitable
anymore when the temperature threshold for solid-liquid phase
transition is reached. The results we provide should therefore be
regarded as qualitative evolutionary trends rather than a quantitative
description of Orcus' thermal evolution. Accounting for sources of
substantial cooling such as convection movements would probably induce
moderate or even low temperature at the present time. Moreover, we
considered a water ice and dust mixture which contains a high fraction
of dust, which in turn induces a high heating power by radiogenic
elements. This dust fraction might actually be lower, and some thermal
properties such as the thermal conductivity might also be reduced by a
large factor in reality. However, even when considering low heating,
our simulations show that Orcus should have experienced a
cryovolcanism episode in its life, especially if ammonia is present in
its bulk composition.
\\

We did not take into account the possible contribution from the
satellite in our simulations, as it is currently not possible to
constrain it. However, if the binary system was created by a giant
impact, we can expect that a large fraction of the main body's surface
layers was removed, thus exposing interior material. Furthermore, the
heat from the impact should contribute to a temporal increasing of
Orcus' upper layers temperature. Tidal heating was not accounted for
either, and might play a role in Orcus thermal evolution if the
satellite is found to be large enough \citep[current rough estimates
from][give a size of the satellite of 1/3 to half the
primary]{Brown+10}.

\subsubsection{A past cryovolcanic event}

Under the effect of an efficient heating due to radiogenic elements, a
liquid phase (water+ammonia, if ammonia is present) is produced, and
could have reached layers close to the surface, eventually freezing,
and/or even propagating through cracks within the ice matrix, thus
erupting at the surface \citep{Stevenson+82}. This process could be
even more efficient with methane, which can trigger explosive
cryovolcanism \citep{Kargel+92, Kargel+95}. Our spectral data show a
compatibility with traces of ammonia and methane at the surface of
Orcus, which means that the interior of the object most probably
contains ammonia. There is nonetheless no easy way to infer the
internal abundance of ammonia from the surface composition, since the
latter would not reflect the internal composition, due to the chemical
modifications induced by space weathering processes. As described by
\cite{Desch+09}, the presence of only a few percent of ammonia in the
interior (by weight, relative to water) can efficiently lower the
viscosity of the ammonia-water mixture once is has melted. Please note
however that large amounts of ammonia might be incompatible with a
high dust/ice ratio such as the one used on our model
\citep{Wong+07}. Our simulations show that we cannot rule out a
cryovolcanic episode in Orcus' past life, which probably supplied most
of the observed crystalline water ice of the surface. Surface erosion
with impact cratering could also contribute to bring to the surface
some fresh material, including crystalline water ice, ammonia, and
methane, refrozen in subsurface layers, but to a much lesser extent
\citep[as computed by][for Charon]{Cook+07}.

\subsubsection{Origin and surviving of  crystalline water ice}

The initial state of water ice in the Solar System is still a matter
of debate:

\begin{itemize}
\item If water ice was initially crystalline when Orcus formed, then
  it could have stayed so over the age of the Solar System in the
  object's interior. Space weathering processes would have induced its
  progressive amorphization at the surface (partial, as we will see
  below, or complete).

\item If water ice was initially amorphous when Orcus formed (as we
  assumed), then our simulations show that it would have been rapidly
  crystallized compared to the age of the Solar System for most of the
  objects' volume. A layer of pristine amorphous ice (a few kilometers
  thick in the canonical case) would have been maintained at the
  surface. In this case, the crystalline water ice would only appear
  at the surface if a supply mechanism was at work at some point of
  the history, or if the erosion by impacts removed the amorphous ice
  over few kilometers \citep[which is currently believed to be an
  inefficient process to explain the detectable quantities of
  crystalline water ice on Orcus, other KBOs and giant planets'
  satellites, as computed by][for Charon]{Cook+07}. However, a more
  detailed modeling of the impact environment of Orcus needs to be
  conducted before completely ruling out the resurfacing effects of
  impacts.
\end{itemize}

In our particular case, the detection of large amounts of crystalline
water ice at the surface of Orcus could indicate that heating from
radiogenic decay and the crystallization of amorphous ice had been an
efficient process, thus bringing the crystallization front close to
the surface, as shown by our thermal evolution model. But most
importantly, our results show that Orcus' thermal history is
compatible with a past cryovolcanic event, which is an efficient
supply mechanism of fresh ices to the surface. We can therefore
envision that most of the crystalline water ice we currently observe
are the actual leftovers from such a geological event. Indeed, recent
laboratory experiments by \cite{Zheng+09}, show that the 1.65$\mu$m
crystalline water ice absorption band is able to survive ion
irradiation doses up to 160$\pm$30 eV for T$\ge$30K. This phenomenon
is temperature-dependent, and becomes even more efficient at higher
temperatures. This means that the amorphization of the crystalline ice
is not complete, and starting at 50K, there is a balance dominated by
thermal recrystallization. From the irradiation doses determined by
\cite{Cooper+03} for the optical thickness of 0.1 mm (\textit{i.e.}
the layers probed by near infrared spectroscopy), Zheng et
al. computed for a KBO, that the 1.65$\mu$m band should be able to
survive during about 6 billion yrs, \textit{e.g.} longer than the
Solar System age. If this result is confirmed, it means that the
40--50K surface of Orcus still holds traces of a past cryovolcanic
event in the form of detectable crystalline water ice, and that
additional resurfacing processes are not necessarily needed.

\section{Summary}

We provided new 0.4--2.35 $\mu$m reflectance spectra and near-infrared
photometry of Plutino (90482) Orcus. Our visible spectrum is
featureless (within the SNR of the data) with a neutral slope
($\sim$2\%/100nm). These results are consistent with previously
published works and point (within the noise level) to a homogeneous
surface at least in the optical. Our near-infrared spectra confirm the
presence of water ice in its crystalline state (thanks to the
1.65$\mu$m absorption band) at the surface of Orcus, as well as the
presence of a 2.2$\mu$m feature, which cannot be readily identified
owing to both the detection level and the multiple candidates. This
band might show a double structure, which deserves to be further
investigated from significantly higher SNR studies than currently
available. These general spectral properties are shared with Pluto's
satellite Charon and Plutino (208996) 2003~AZ$_{84}$.
\\

Radiative transfer Hapke models show that a simple mixture of water
ice (amorphous and crystalline) is not sufficient to properly describe
the data in the H and K band. Orcus therefore does not have a
Haumea-type surface, although H$_2$O should be the dominant
ice. Additional absorbents should be invoked, which may be volatiles
and their possible irradiation products. Further Hapke modeling shows
that the 2.2$\mu$m region and overall spectrum from 0.4 to 2.35 $\mu$m
are best described by an intimate mixture of ammonium (NH$_4^+$) and
traces of ethane (C$_2$H$_6$), which are most probably solar
irradiation products, in addition to water ice (mostly in its
crystalline state). The presence of ammonium even helps to better
reproduce the visible spectrum, an effect which was not expected
\textit{a priori}. We find also that our data are compatible with
Hapke models of an equivalent (slightly lower) statistical level
including a mixture of water ice, methane, and ammonia (pure of
diluted); the latter are the parent species of ethane and
ammonium. Methane is particularly interesting to reproduce the blue
slope longward $\sim$2.3$\mu$m, which might also be the shortward
wing of an additional absorption feature. Focused studies should be
conducted in close connection with laboratory investigations to better
understand the methane and ammonia chemistry.
\\

In this work, we showed that the Orcus spectrum could be compatible
with the presence of volatile ices on its surface \citep[as also
hypothesized by][]{Barucci+08}, and their involatile irradiation
products (first invoked here). Our findings merits confirmation by
higher SNR data, and we believe that Plutino (208996) 2003~AZ$_{84}$
and Orcus deserve joint detailed studies, to be also connected with
Charon properties, as they share common spectral
properties. \cite{Schaller+07} developed a simple model of atmospheric
escape to predict how a given KBO can preserve a primordial surface
inventory of CO, N$_2$ and CH$_4$. Orcus, Charon and Quaoar are
located in a relative narrow region in their Fig. 1, showing the
equivalent temperature (K) as a function of the object's diameter
(km). Orcus is located in the ``all volatiles lost'' region, where
methane and nitrogen ices in particular should have completely escaped
from the surface. That our Orcus data could be compatible with the
presence of volatiles (here CH$_4$) would imply that (i) such an
atmospheric model is too simple to describe the actual volatile
content of a particular object (as it was emphasized by the authors in
their conclusions); (ii) Orcus might be larger than the size we
currently know from far-IR data (and \textit{Herschel Space
  Observatory} observations might allow us to answer this question,
although from the Schaller and Brown model, a factor of $\sim$1.7
would be needed); (iii) there is a supply of volatile ices from the
interior of the object; (iv) the actual situation is a combination of
the previous points.
\\

We investigated the thermal properties of Orcus over its lifetime to
understand how crystalline water ice can be produced and test if
geological events such as cryovolcanism can be possible and provide
fresh ices to the surface. We used a 3D thermal evolution model by
\cite{Guilbert+10} that simulates the temperature of the whole object
as a function of its size, density, porosity, orbit, formation delay
and radiogenic elements primordial inventory ($^{26}$Al, $^{60}$Fe,
$^{53}$Mn, $^{40}$K, $^{232}$Th, $^{235}$U, $^{238}$U). The bulk
assumed composition is dust (77\%) plus amorphous water ice (23\%),
the relative mass fractions being derived from the input 10\%
porosity. Various initial configurations have been simulated with
different formation delays (from 1 to over 10 Myr) including different
amounts of short-lived radiogenic elements (with the case no
short-lived elements at all) to evaluate the differentiation state of
Orcus. Our simulations show that a layer of amorphous water ice of a
few kilometers thick would be maintained at the surface of the body,
while the remaining of the volume is occupied by crystalline water
ice. For all cases we considered, the heating induced by radiogenic
elements leads to the melting of the core. This effect is strengthened
if ammonia is present inside the body, because it reduces the melting
point temperature of the mixture and allows for the production of a
liquid phase even under moderate heating. This liquid phase could
easily reach layers close to the surface through cracks in the ice
matrix (if not directly produced in subsurface layers in the more
extreme cases), and therefore make Orcus and intermediate to large
water-bearing KBOs the subject of potential astrobiological interest.
\\

As a consequence, even if it is premature to suggest that Orcus did
undergo a \textit{recent} event of cryovolcanism, our results show
that this event did take place at some point of Orcus' history and
most probably supplied crystalline water ice to the surface. Recent
laboratory experiments by \cite{Zheng+09} show that the
ion-irradiation-induced amorphization of crystalline water ice on a
KBO at $\sim$40K can be incomplete and be efficiently balanced by
re-crystallisation (starting at 50K), so that crystalline water ice
can survive and still be detectable at the surface over timescales
longer than the Solar System age. Both from the latter results and our
model, we can envision that some of the crystalline ice we currently
observe at the surface of Orcus is actually the remnant of a past
cryovolcanic event. In addition, in the most probable case of our
simulations, a large volume of volatiles are currently present in
Orcus subsurface layer, awaiting to be revealed by future surface
alterations.

\begin{acknowledgements}

  We thank D. Jewitt for fruitful suggestions and are grateful to
  J.C. Cook for kindly sharing with us estimated optical constants of
  ammonium, to S. Fornasier et al. for providing us with their Orcus
  visible spectra, and to S. Spjuth for the computation of the visible
  geometric albedo of Charon. This material is based upon work
  supported by the National Aeronautics and Space Administration
  through the NASA Astrobiology Institute under Cooperative Agreement
  No. NNA04CC08A issued through the Office of Space Science. B. Yang
  acknowledges support from a NASA Origins grant and
  A. Guilbert-Lepoutre from a NASA Herschel grant, both to
  D. Jewitt. Part of the data presented here were obtained at the
  W.M. Keck Observatory, which is operated as a scientific partnership
  among the California Institute of Technology, the University of
  California and NASA. Based on observations obtained at the Gemini
  Observatory, which is operated by the Association of Universities
  for Research in Astronomy, Inc., under a cooperative agreement with
  the NSF on behalf of the Gemini partnership. The authors wish to
  recognize and acknowledge the very significant cultural role and
  reverence that the summit of Mauna Kea has always had within the
  indigenous Hawaiian community.

\end{acknowledgements}

%%__________________________________________________________________

\bibliographystyle{aa}
\bibliography{mnemonic,14296bib}

\end{document}